\documentclass[conference]{IEEEtran}
\IEEEoverridecommandlockouts

\usepackage{cite}
\usepackage{amsmath,amssymb,amsfonts}
\usepackage{algorithmic}
\usepackage{graphicx}
\usepackage{textcomp}
\usepackage{xcolor}
\usepackage{cite}
\usepackage{subcaption}
\def\BibTeX{{\rm B\kern-.05em{\sc i\kern-.025em b}\kern-.08em
    T\kern-.1667em\lower.7ex\hbox{E}\kern-.125emX}}

\newtheorem{thm}{Theorem}[section] 

\newtheorem{prop}{Proposition}[section]

\newcommand{\sign}{\mathrm{sign}}

\newcommand{\cA}{\mathcal{A}}

\newcommand{\cU}{\mathcal{U}}

\newcommand{\cP}{\mathcal{P}}

\newcommand{\bd}{{\boldsymbol{d}}}

\newcommand{\bP}{\boldsymbol{P}}
\newcommand{\bT}{\boldsymbol{T}}

\newcommand{\bx}{\boldsymbol{x}}
\newcommand{\bt}{\boldsymbol{t}}

\newcommand{\bZ}{\boldsymbol{Z}}

\newcommand{\be}{\boldsymbol{e}}
\newcommand{\ba}{\boldsymbol{a}}
\newcommand{\bb}{\boldsymbol{b}}

\newcommand{\bq}{\boldsymbol{q}}

\newcommand{\bz}{\boldsymbol{z}}

\begin{document}

\title{Gridless Chirp Parameter Retrieval via Constrained Two-Dimensional Atomic Norm Minimization\\
\thanks{D. Yang and F. Xi contributed equally to this work. D. Yang was supported in part by the Research Development Fund (RDF-23-02-080) from Xi'an Jiaotong-Liverpool University. F. Xi was supported in part by the National Natural Science Foundation of China under grant No. 62471230.}
}

\author{\IEEEauthorblockN{Dehui Yang}
\IEEEauthorblockA{\textit{School of Mathematics and Physics} \\
\textit{Xi'an Jiaotong-Liverpool University}\\
Suzhou, China 215123\\
Email: dehui.yang@xjtlu.edu.cn}
\and
\IEEEauthorblockN{Feng Xi}
\IEEEauthorblockA{\textit{Department of Electronic Engineering} \\
\textit{Nanjing University of Science and Technology}\\
Nanjing, China 210094\\
Email: xifeng@njust.edu.cn}
}

\maketitle

\begin{abstract}
This paper is concerned with the fundamental problem of estimating chirp parameters from a mixture of linear chirp signals. Unlike most previous methods, which solve the problem by discretizing the parameter space and then estimating the chirp parameters, we propose a gridless approach by reformulating the inverse problem as a constrained two-dimensional atomic norm minimization from structured measurements. This reformulation enables the direct estimation of continuous-valued parameters without discretization, thereby resolving the issue of basis mismatch. An approximate semidefinite programming (SDP) is employed to solve the proposed convex program. Additionally, a dual polynomial is constructed to certify the optimality of the atomic decomposition. 
Numerical simulations demonstrate that exact recovery of chirp parameters is achievable using the proposed atomic norm minimization. 
\end{abstract}


\section{Introduction}
\subsection{Motivation}
Chirp signals naturally arise in a variety of applications in signal processing, such as radar, sonar, wireless communications, speech, and laser systems. They are commonly used to model non-stationary signals, where the frequency content varies over time. For instance, in wireless communications, a well-known technique called chirp spread spectrum employs wideband linear frequency modulated chirps to encode information \cite{spread-specturm}. Spectrum analysis in this case involves estimating the unknown initial frequency and frequency rate parameters from sampled measurement data. In radar applications, chirps are used to estimate the trajectories of moving objects, making the estimation of these unknown parameters crucial for tracking and localization of targets \cite{radarbook}. Another example arises in musical audio, such as vibrato of a singing voice and attack of a plucked string, where the signal often exhibits time-varying features and can be modeled as linear chirp sinusoids \cite{julian}. Estimating the parameters of chirp sinusoids from noisy data can significantly enhance the quality of signal re-synthesis. 

In this paper, we focus on a specific class of chirp signals: linear chirps. Linear chirps are a subset of a broader class of signals with polynomial phases. Specifically, a linear chirp is a polynomial phase signal of degree two, parameterized by its initial frequency and linear frequency rate, in addition to the unknown amplitude.  Our goal is to estimate the number of linear chirp components, as well as each component's amplitude, initial frequency, and frequency rate, from discrete-time samples of their mixtures.

\subsection{Related Work and Contributions}
Various algorithms have been developed over the past decades for the effective estimation of chirp parameters in the literature. Here, we review the methods most related to our work. In \cite{djuric}, a method based on time-domain maximum likelihood estimation is introduced for the single linear chirp case. In \cite{doweck-1}, the maximum likelihood estimation-based approach is used to estimate the fundamental initial frequency and frequency rate of linear chirp signals with harmonic components and random amplitudes. In \cite{volcker}, a subspace-based approach that exploits  the structure of the sample covariance matrix is introduced, and a rank reduction technique is developed to estimate the chirp parameters. Another line of related work has exploited sparse recovery techniques for parameter estimation of chirp signals \cite{sward}. However, this approach relies on discretization and thus suffers from the risk of basis mismatch, a common issue in the estimation of continuous-valued parameters. More recently, \cite{tf-atomic} proposes an atomic norm minimization approach for sparse time-frequency representation, which can be used for gridless chirp parameter estimation. However, their approach does not exploit the inherent structure of linear chirps and does not directly estimate the chirp parameters. 

The main contribution of this work is the formulation of the gridless chirp parameter retrieval problem as sparse recovery from structured measurements using two-dimensional atomic norm minimization. To address the inherent ambiguity and ill-posed nature of the inverse problem, a constraint on the chirp frequency rate parameter is introduced. An approximate SDP formulation is then derived as a computationally feasible solver for the proposed atomic norm minimization.
Furthermore, we provide a dual certificate condition to certify the optimal atomic decomposition by constructing a dual polynomial. Due to the high computational complexity of the proposed SDP, a faster alternative based on a decoupled two-dimensional SDP is designed for the numerical experiments. 
Numerical simulations demonstrate that the proposed method is effective for estimating the chirp frequency and frequency rate from measurement data.

\section{Signal Model and Problem Formulation}
\subsection{Signal Model}\label{signal_model}
Consider a mixture of $K$ linear chirp signals
\begin{equation}
\label{eq:dt-chirp}
\bx(n) = \sum_{k = 1}^K c_k e^{j 2\pi (f_k + \theta_k n)n}, ~~n = 0, \cdots, N-1,
\end{equation}
where $\bx(n) \in \mathbb{C}$ are samples of the observation data, $\left\{c_k\right\} \subset \mathbb{C}, \left\{f_k\right\} \subset [0, 1)$, and $\left\{\theta_k\right\} \subset [0, 1)$ are the unknown continuous-valued parameters associated with the complex exponentials $\left\{c_k e^{j 2\pi (f_k + \theta_k n)n}\right\}$. Here, $f_k$ and $\theta_k$ represent the {\em normalized initial frequency} and the {\em normalized frequency rate} of the chirp signal $c_k e^{j 2\pi (f_k + \theta_k n)n}$, respectively. Our goal is to reliably recover the unknown parameters $\left\{(c_k, f_k, \theta_k)\right\}$ from $\bx(n)$. 

The signal model (\ref{eq:dt-chirp}) generalizes narrowband signals in line spectrum estimation to wideband signals.  Compared to the complex exponentials $\left\{c_k e^{j 2\pi f_k n}\right\}$ that are encountered in line spectrum estimation, the linear chirp signals have additional parameters $\left\{ \theta_k \right\}$, which characterize the linearly time-varying behavior of the unknown frequency content over time. It is worth noting that when $\left\{\theta_k\right\}$ are all zeros, the problem reduces to frequency estimation in line spectrum estimation \cite{tang2013compressed}.

One may obtain the discrete-time signal model (\ref{eq:dt-chirp}) from a band-limited continuous-time signal $x(t)$ defined over the time interval $[0, T]$, i.e., 
\begin{equation}
\label{eq:ct-chirp}
x(t) =  \sum_{k = 1}^K c_k e^{j 2\pi (F_k + \Theta_k t)t},~~~0 \leq t \leq T,
\end{equation}
where $F_k$ and $\Theta_k$\footnote{ For simplicity, we assume the frequency rate $\Theta_k > 0$ and the frequency increases over time} are the analog initial frequency and frequency rate, respectively. Then the bandwidth of the $k$th chirp component $c_k e^{j 2\pi (F_k + \Theta_k t)t}$ is given by $B_k = T \Theta_k$. Assume that all of the analog frequencies of $x(t)$ are bandlimited to $[-W, W]$, and denote $F_s$ as the sampling rate, which satisfies the Nyquist sampling theorem with $F_s \geq 2W$.
By taking $N$ regularly spaced samples of $x(t)$ at times  $\left\{\frac{0}{F_s}, \frac{1}{F_s}, \cdots, \frac{N-1}{F_s}\right\}$, we have
\begin{equation}
\begin{aligned}
\label{eq:sampling}
x\left(t=\frac{n}{F_s}\right) & = \sum_{k = 1}^K c_k e^{j 2\pi \left(\frac{F_k}{F_s} + \frac{\Theta_k}{F_s^2} n\right)n}, n=0, 1, \cdots, N-1.
\end{aligned}
\end{equation}
Redefining $\bx(n) = x(n/F_s)$, $f_k = F_k / F_s$, and $\theta_k = \Theta_k / F_s^2$, we obtain a discrete-time signal model with normalized frequency in the interval of $[-\frac{1}{2}, \frac{1}{2}]$, which is the signal model (\ref{eq:dt-chirp}) after a trivial translation in the frequency domain.

\subsection{Problem Formulation}
In this section, we formulate the parameter estimation of linear chirps as the super-resolution of a two-dimensional line spectrum estimation problem from structured measurement data. 

To start with, 
define $M = (N-1)^2 + 1$, $J = \left\{0, 1, \cdots, N-1\right\} \times \left\{0, 1, \cdots, M-1 \right\}$, and 
denote 
\begin{equation}
\ba(f)  = \begin{bmatrix} e^{j2\pi 0 f} & e^{j2\pi 1 f} & \cdots & e^{j2\pi (N-1)f} \end{bmatrix}^T \in \mathbb{C}^{N\times 1}
\end{equation}
and 
\begin{equation}
\bb(\theta)  = \begin{bmatrix} e^{j2\pi 0 \theta} & e^{j2\pi 1 \theta} & \cdots & e^{j2\pi (N-1)^2 \theta} \end{bmatrix}^T \in \mathbb{C}^{M\times 1}.
\end{equation}
In addition, we define $\bd(f, \theta)$ as the Kronecker product of $\ba(f)$ and $\bb(\theta)$, i.e.,  
\begin{equation}
\begin{aligned}
\bd(f, \theta) & = \ba(f) \otimes \bb(\theta) \in \mathbb{C}^{NM \times 1}.
\end{aligned}
\end{equation} 
Denote $\bz^{\star} \in \mathbb{C}^{NM \times 1}$ as a mixture of $K$ components from the set $\left\{\bd(f_k, \theta_k),~f_k\in [0, 1),\theta_k \in [0, 1), k = 1, \cdots, K\right\}$, i.e., 
\begin{equation}
\begin{aligned}
\bz^{\star} & = \sum_{k = 1}^K  c_k\ba(f_k) \otimes \bb(\theta_k)  = \sum_{k = 1}^K c_k \bd(f_k, \theta_k).
\end{aligned}
\end{equation}
For the $n$th measurement $\bx(n)$ in (\ref{eq:dt-chirp}), we can write 
\begin{equation}
\begin{aligned}
\bx(n) & = \sum_{k = 1}^K c_k e^{j 2\pi (f_k + \theta_k n)n} \\
& = \sum_{k = 1}^K  c_k \underbrace{ \ba(f_k) (n)  \cdot \bb(\theta_k) (n^2)}_{\bd(f_k, \theta_k)(nM + n^2)},
\end{aligned}
\end{equation}
where we use the notations $\ba(f_k) (n)$, $\bb(\theta_k) (n^2)$, and $\bd(f_k, \theta_k)(nM + n^2)$ to represent the $n$th, $n^2$th, and $(n M + n^2)$th elements of $\ba(f_k)$, $\bb(\theta_k)$, and $\bd(f_k, \theta_k)$, respectively. \\
Denote $\be_{nM + n^2}$ as the $(nM + n^2)$th\footnote{For convenience, here we have used zero-based index.} column of an $NM \times NM$ identity matrix. Then one can write $\bx$ as a collection of structured linear measurements of $\bz^{\star}$, i.e., 
\begin{equation}
\begin{aligned}
\label{eq:reformaution}
\bx  & = \cP\left(\sum_{k = 1}^K c_k \bd(f_k, \theta_k)\right) = \cP(\bz^{\star}),
\end{aligned}
\end{equation} 
where the linear operator $\cP: \mathbb{C}^{NM} \rightarrow \mathbb{C}^N$ is defined as $\bx(n) = \left\langle \bz^{\star}, \be_{nM + n^2}\right  \rangle =  \be_{nM + n^2}^H \bz^{\star},~n = 0, 1, \cdots, N-1$. Note that the linear operator $\cP(\cdot)$ can be written as $\cP(\bz^{\star}) = \bP\bz^{\star}$, where $\bP$ is an $N\times NM$ measurement matrix with each row being $\be_{nM + n^2}^H,~n = 0, 1, \cdots, N-1$. Therefore, estimating the unknown parameters of the chirp signal is equivalent to recovering $\bz^{\star}$ and its associated parameters $\left\{(c_k, f_k, \theta_k)\right\}$ from $\bx(n), n = 0, 1, \cdots, N-1$ in (\ref{eq:reformaution}). 

The reformulation (\ref{eq:reformaution}) is an instance of continuous-valued two-dimensional line spectrum estimation from compressive measurements \cite{chi-2d}. In contrast to the conventional random sampling paradigm used in the framework of compressive sensing, our measurement scheme is derived from the chirp signal model (\ref{eq:dt-chirp}) and is therefore highly structured. 

\subsection{Identifiability and Aliasing}
It is easy to verify that for each chirp component $c_k e^{j 2\pi (f_k + \theta_k n)n}$ in (\ref{eq:dt-chirp}), in addition to the underlying true parameters $\left\{(f_k, \theta_k)\right\}$, there exists another set of parameters $\left\{(\bar{f}_k, \bar{\theta}_k)\right\} \subset [0, 1) \times [0, 1)$ with $|f_k - \bar{f}_k| = \frac{1}{2}$ and $|\theta_k - \bar{\theta}_k| = \frac{1}{2}$, which satisfies $c_k e^{j 2\pi (f_k + \theta_k n)n} = c_k e^{j 2\pi (\bar{f}_k + \bar{\theta}_k n)n}$. Due to this ambiguity, the parameter set $\left\{f_k, \theta_k\right\}$ is generally not recoverable if we search for the unknown parameters $f_k$ and $\theta_k$ within the unit interval $[0, 1)$. 

To prevent aliasing, in Section \ref{signal_model} we have shown that $F_s \geq 2W$. For $x(t)$ over the time interval $[0, T]$ as defined in (\ref{eq:ct-chirp}),  we have $\max_k B_k = T \max_k \Theta_k = \frac{N}{F_s} \max_k \Theta_k$. Since $B_k$ is bounded by $2W$, we obtain $F_s \geq  \frac{N}{F_s} \max_k \Theta_k$, which is equivalent to $\max_k \theta_k = \max_k \frac{\Theta_k}{F_s^2} \leq \frac{1}{N}$. This shows that, under the Nyquist sampling rate, $\theta_k$ should be reasonably small and is bounded by $\frac{1}{N}$. By integrating this prior knowledge of $\theta_k$ into the development of the atomic norm minimization framework, the ill-posed inverse problem (\ref{eq:reformaution}) becomes more tractable.

\section{Constrained Two-Dimensional Atomic \\ Norm Minimization} \label{Section:Constrained Atomic Norm Minimization}
\subsection{Constrained Atomic Norm Minimization and Its SDP Representation}
The atomic norm, first proposed in \cite{chandrasekaran2012convex}, provides a general framework for enforcing sparsity in a signal or dataset that is a superposition of a few atoms from a dictionary \cite{tang2013compressed,chi-2d,dyang}. 
To address the ambiguity and aliasing issues discussed above, we propose solving a constrained atomic norm minimization problem by imposing a constraint on $\theta_k$. Specifically, we choose the parameters $\left\{\theta_k\right\}$ from a subset $[0, \cU] \subset [0, 1)$ with $\cU \leq \frac{1}{N}$. Motivated by this, we define the constrained atomic set as 
\begin{equation}
\cA _c= \left\{ \bd (f, \theta) | f \in [0, 1), \theta \in [0, \cU]\right\},
\end{equation}
where $\cU$ is an upper bound for the $\theta$ parameter, reflecting the prior knowledge of the normalized frequency rate. The corresponding constrained atomic norm is then defined as 
\begin{equation}
\label{constrained-atomic-norm}
\begin{aligned}
|| \bz ||_{\cA_c} & = \inf\left\{t> 0: ~\bz \in t \text{conv}(\cA_c)\right\} \\
& = \inf_{c_i, f_i, \theta_i \in [0, \cU]} \left\{ \sum_{i = 1} |c_i|:~\bz = \sum_i c_i \bd(f_i, \theta_i) \right\}
\end{aligned}
\end{equation}
To enforce sparsity in the constrained atomic representation, we solve 
\begin{equation}
\label{constrained-atomic-minimization}
\begin{aligned}
& \min_{\bz}~||\bz||_{\cA_c}~~~~\text{subject~to}~~\bx = \cP(\bz)
\end{aligned}
\end{equation}  

The constrained atomic norm $||\bz||_{\cA_c}$, induced by the atomic set $\cA_c$, is related to prior work on frequency-selective Vandermonde decomposition of Toeplitz matrices in one-dimensional (1-D) \cite{mishra,yangzai-1} and multi-dimensional (M-D) cases \cite{li-yc}. It has been shown that $||\bz||_{\cA_c}$ admits an equivalent SDP formulation in the 1-D case and an approximate SDP formulation  in the M-D case. Before presenting the approximate SDP characterization for $||\bz||_{\cA_c}$, we first introduce the Hermitian trigonometric polynomials and two-fold Toeplitz matrices, which are necessary for the discussion. \\
{\em I. Hermitian Trigonometric Polynomials}: A Hermitian trigonometric polynomial of degree one is defined as 
\begin{equation}
\label{degree-one-poly}
g(z) = r_1 z^{-1} + r_0 + r_{-1} z,~~r_{-1} = \bar{r}_1, r_0 \in \mathbb{R}, 
\end{equation}
where $z$ is a complex argument and $\bar{\cdot}$ denotes the complex conjugate operation.
When $g(z)$ is evaluated on the unit circle, i.e., $z = e^{j 2\pi f}$, we have 
\begin{equation}
g(e^{j2\pi f}) = r_1 e^{-j2\pi f} + r_0 + r_{-1} e^{j2\pi f} = r_0 + 2\text{Re}(r_1e^{-j 2\pi f}),
\end{equation}
which is real-valued.\\
{\em II. Two-Fold Toeplitz Matrix (a.k.a. Doubly Toeplitz Matrix)}: Given a $(2N -1)\times (2M-1)$ matrix $\bT$, the two-fold Toeplitz matrices $\text{Toep}(\bT)\footnote{Hereafter, we use the notation Toep($\cdot$) to represent a two-fold Toeplitz matrix generated from the underlying matrix.} \in \mathbb{C}^{NM \times NM}$ and $\text{Toep}(\bT^g) \in \mathbb{C}^{N(M-1) \times N(M-1)}$ are defined as 
\begin{equation}
\begin{aligned}
\label{tf-toep-outer}
& \text{Toep}(\bT) = \begin{bmatrix}
\bT_0 & \bT_{-1} & \cdots & \bT_{-(N-1)} \\
\bT_1 & \bT_{0} & \cdots & \bT_{-(N - 2)} \\
\vdots & \vdots & \ddots & \vdots \\
\bT_{N - 1} & \bT_{N -2 } & \cdots & \bT_0 
\end{bmatrix}, \\
& \text{Toep}(\bT^g) = \begin{bmatrix}
\bT_0^g & \bT_{-1}^g & \cdots & \bT_{-(N-1)}^g \\
\bT_1^g & \bT_{0}^g & \cdots & \bT_{-(N - 2)}^g \\
\vdots & \vdots & \ddots & \vdots \\
\bT_{N - 1}^g & \bT_{N -2}^g & \cdots & \bT_0^g 
\end{bmatrix},
\end{aligned}
\end{equation}
where each $\bT_{\ell}$ in $\text{Toep}(\bT)$ is an $M \times M$ Toeplitz matrix formed by the $\ell$th row of $\bT$
with $\bT_{\ell}[i, j] = \bT[\ell, i-j],~0\leq i, j\leq M-1$, and $\bT_{\ell}^g$, is an $(M-1) \times (M-1)$ Toeplitz matrix with each entry 
\begin{equation}
\label{tf-toep-g-inner}
\bT_{\ell}^g[i', j'] = r_1 \bT[\ell, i'-j' + 1] + r_0 \bT[\ell, i'-j'] + r_{-1}\bT[\ell, i'-j' - 1],
\end{equation}
where $0\leq i', j'\leq M-2$ and $g$ is a degree one trigonometric polynomial defined in (\ref{degree-one-poly}).

 Based on prior works \cite{yangzai-1} and \cite{li-yc}, we have the following proposition, which is a special case of Proposition 2 in \cite{li-yc} and provides an approximate SDP formulation for the constrained atomic norm  (\ref{constrained-atomic-norm}).
 \begin{prop}
 \label{appro-SDP}
 The constrained atomic norm (\ref{constrained-atomic-norm}) is lower bounded by the minimum value of the SDP shown below 
 \begin{equation}
 \begin{aligned}
 \label{eq:approx-SDP}
 & ||\bz||_{\cA_c} \geq \min_{\bT, t}~ \frac{1}{2|J|}{\text {trace}}(\text{Toep}(\bT)) + \frac{1}{2}t\\
 & {\text{subject~to}}~~\begin{bmatrix}
 \text{Toep}(\bT) & \bz \\
 \bz^H & t
\end{bmatrix}   \succeq 0,~\text{Toep}(\bT^{g})   \succeq 0,
\end{aligned}
 \end{equation}
 where $g$ is defined in (\ref{degree-one-poly})
 with $r_0 = -2\cos(\cU \pi)$ and $r_1 = e^{j\pi \cU}$, and $|J|$ is the cardinality of $J$. The Toeplitz matrices $\text{Toep}(\bT)$ and $\text{Toep}(\bT^g)$ are defined according to (\ref{tf-toep-outer}). Furthermore, if $\text{rank}(\text{Toep}(\bT)) < N$, the inequality in (\ref{eq:approx-SDP}) becomes an equality. In this case, the SDP characterization is an equivalent representation of the constrained atomic norm.
 \end{prop}
 
As a result of this, we can approximate the constrained atomic norm minimization (\ref{constrained-atomic-minimization}) via the following SDP: 
\begin{equation}
\label{primal-SDP}
\begin{aligned}
& \min_{\bT, t, \bz}~\frac{1}{2|J|}{\text {trace}}(\text{Toep}(\bT)) + \frac{1}{2}t  \\
&  {\text{subject~to}}~~\bx = \bP \bz\\
&~~~~~~~~~~~~\begin{bmatrix}
\text{Toep}(\bT) & \bz \\
 \bz^H & t
\end{bmatrix}   \succeq 0,~\text{Toep}(\bT^g)  \succeq 0, 
\end{aligned}
\end{equation}
which is computationally feasible. Off-the-shelf solvers, such as CVX \cite{grant2008cvx}, can be used to solve (\ref{primal-SDP}). It is worth noting that the dual of (\ref{primal-SDP}) is also an SDP. Due to space limitations, we leave the details of the derivation for future work.  

\subsection{Dual Atomic Norm and Dual Certificate}
The dual norm associated with the constrained atomic set $\cA_c$ is given by 
\begin{equation}
\label{dual-constrained-atomic-norm}
||\bq||_{\cA_c}^{*} = \sup_{||\bz||_{\cA_c} \leq 1} \left\langle \bq, \bz\right  \rangle_{\mathbb{R}} = \sup_{f \in [0, 1), \theta \in [0, \cU]} | \left\langle \bq, \bd(f, \theta) \right  \rangle_{\mathbb{R}}|,
\end{equation}
where $\langle \cdot, \cdot  \rangle_{\mathbb{R}}$ denotes the real part of the inner product.
Standard Lagrangian analysis shows that the dual of (\ref{constrained-atomic-minimization}) is given by
\begin{equation}
\label{dual-constrained-atomic-norm-minimization}
\begin{aligned}
\max_{\bq} \left\langle \bq, \bx \right \rangle_{\mathbb{R}}~~~\text{subject~to}~~~||\cP^{*}(\bq)||_{\cA_c}^{*} \leq 1,
\end{aligned}
\end{equation}
where $\cP^{*}$ is the adjoint operator of the linear operator $\cP$. Since the linear operator $\cP$ is characterized by the measurement matrix $\bP$, we have $\cP^*(\bq) = \bP^H \bq \in \mathbb{C}^{NM \times 1}$. 

Define the dual polynomial 
\begin{equation}
Q(f, \theta) = \bd^H(f, \theta) \bP^H \bq,
\end{equation}
which serves to certify the optimal atomic decomposition, as shown in the following theorem.
\begin{thm}
\label{chirp-atomic-norm-dual-certificate}
Suppose that the constrained atomic set $\mathcal{A}_c$ is composed of atoms of the form $\bd(f, \theta)$ with $(f, \theta) \in [0,1) \times [0, \cU]$. Define the set $\mathbb{D} = \left\{(f_k, \theta_k)\right\}_{k=1}^K$. Let $\widehat{\bz}$ be the optimal solution to (\ref{constrained-atomic-minimization}). Then $\widehat{\bz} = \bz^{\star}$ is the unique optimal solution if the following two conditions are satisfied: \\
1) There exists a dual polynomial
\begin{equation}
\label{dual cond}
\begin{aligned}
Q(f, \theta) & = \left \langle \mathcal{P}^{*}(\bq), \bd(f, \theta) \right \rangle\\
& = \sum_{n=0}^{N-1} \bq(n) e^{-j2\pi (fn + \theta n^2)} 
\end{aligned}
\end{equation}
satisfying   
\begin{equation}
\label{dual cond3}
\begin{aligned}
& Q(f_k, \theta_k) = \sign(c_k), ~~~\forall~(f_k, \theta_k) \in \mathbb{D}\\
& |Q(f, \theta)| <1,~~~\forall~(f, \theta) \notin \mathbb{D},
\end{aligned}
\end{equation}
where $\bq$ is a dual optimizer and $\sign(c_k) := \frac{c_k}{|c_k|}$.
\label{optimalitycond}\\
2) $\left\{ \bP \bd(f_k, \theta_k), k = 1, \cdots, K\right\}$ with $K < N$ is a linearly independent set.
\end{thm}

\subsection{Decoupled Two-Dimensional Atomic Norm Minimization}
While solving (\ref{primal-SDP}) is computationally feasible using off-the-shelf solvers, it requires a significant amount of computational resources due to the high dimensionality of the positive semidefinite (PSD) constraints. Specifically, the PSD constraints in (\ref{primal-SDP}) are of size $O(N^3 \times N^3)$. According to \cite{zhe-zhang-decouple}, for each iteration of the interior-point method, the computational cost is $O(N^9)$, which is extremely high even for small values of $N$. To alleviate this, we leverage the decoupled two-dimensional atomic norm formulation for efficient approximation of the two-dimensional atomic norm minimization \cite{zhe-zhang-decouple,feng-decouple}. In particular, the SDP relaxation of the constrained version of the decoupled two-dimensional atomic norm in our constrained case is given by 
\begin{equation}
\label{decouple}
\begin{aligned}
& \min_{\bt_1,\bt_2, \bZ } \frac{1}{2\sqrt{NM}}\left({\text {trace}}(\text{toep}(\bt_1)) + {\text {trace}}(\text{toep}(\bt_2)) \right)  \\
& {\text{subject~to}}~~\bx = \bP\text{vec}(\bZ)\\
& ~~~~~~~~~~~\begin{bmatrix}
 \text{toep}(\bt_1) & \bZ^H \\
 \bZ & \text{toep}(\bt_2)
\end{bmatrix}   \succeq 0,~\text{toep}(\bt_2^g)  \succeq 0, 
\end{aligned}
\end{equation}
where $\bt_1 \in \mathbb{C}^{(2N - 1)\times 1}, \bt_2 \in \mathbb{C}^{(2M-1)\times 1}, \bZ \in \mathbb{C}^{M\times N}$, $\text{toep}(\cdot)$ represents the one-folder Toeplitz matrix formed by the underlying variable, and the Toeplitz matrix $\text{toep}(\bt^g)$ is formed in a similar way as $\text{Toep}(\bT^g)$ in (\ref{primal-SDP}). 
It is easy to see that the PSD constraint in (\ref{decouple}) is of size $O(N^2 \times N^2)$, which is an order of $N$ smaller than that in (\ref{primal-SDP}). In the numerical experiments section, we implement the constrained version of the decoupled two-dimensional atomic norm minimization for solving (\ref{primal-SDP}) using CVX.

\section{Numerical Experiments}
In this section, we present numerical simulations to demonstrate the effectiveness of the proposed approach for chirp parameter estimation. For demonstration purposes, we set $N = 25$ and generate $K=2$ pairs of chirp parameters: $(f_1, \theta_1) = (0.165, 0.013)$ and $(f_2, \theta_2) = (0.524, 0.0075)$. The amplitudes of the chirp signals are both set to $1$. We constrain the upper bound of the normalized frequency rate $\theta$ to the interval $[0, 0.02]$, i.e., $\cU = 0.02$. The locations of the ground truth parameters are depicted in Fig. 1(a). Fig. 1(b) shows the recovered parameters using the discretized sparse recovery method of basis pursuit. It is implemented with a discretization resolution of $0.01$ for $f$ and $0.001$ for $\theta$. Fig. 1(c) illustrates the recovered parameters using the proposed atomic norm minimization, superimposed on the ground truth. It is observed that the parameters are perfectly recovered with the noise-free measurements. To evaluate the performance of the proposed approach under noisy conditions, we add additive white Gaussian noise generated from a normal distribution to the measurements. The signal-to-noise ratio (SNR) is set to $20$dB. Fig. 1(d) shows the recovered parameters along with the ground truth. Our proposed approach is able to reliably recover the chirp parameters even with noisy measurements. 

For the noise-free exact recovery case as shown in Fig. 1(c), we also examine the corresponding dual polynomial, which is formed by plugging the optimal dual variable associated with the equality constraint of the decoupled two-dimensional atomic norm minimization into (\ref{dual cond}). Fig. 2 plots the absolute value of the dual polynomial along with the ground truth. It is observed that the dual polynomial achieves its peak only at the locations of the ground truth, which is consistent with Theorem \ref{chirp-atomic-norm-dual-certificate}.

\begin{figure}[htb]
   \begin{subfigure}[b]{.49\columnwidth}
    \includegraphics[width=\columnwidth, height=1.5in]{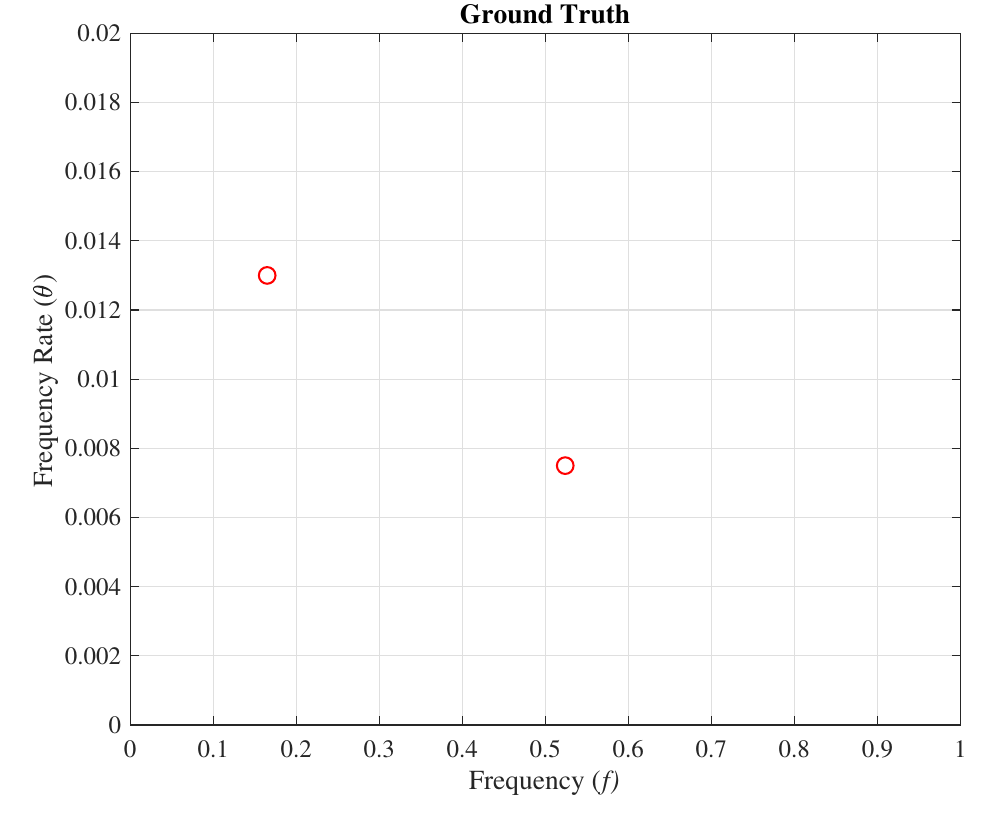}
     \caption{}
   \end{subfigure} 
   \begin{subfigure}[b]{.49\columnwidth}
    \includegraphics[width=\columnwidth, height=1.5in]{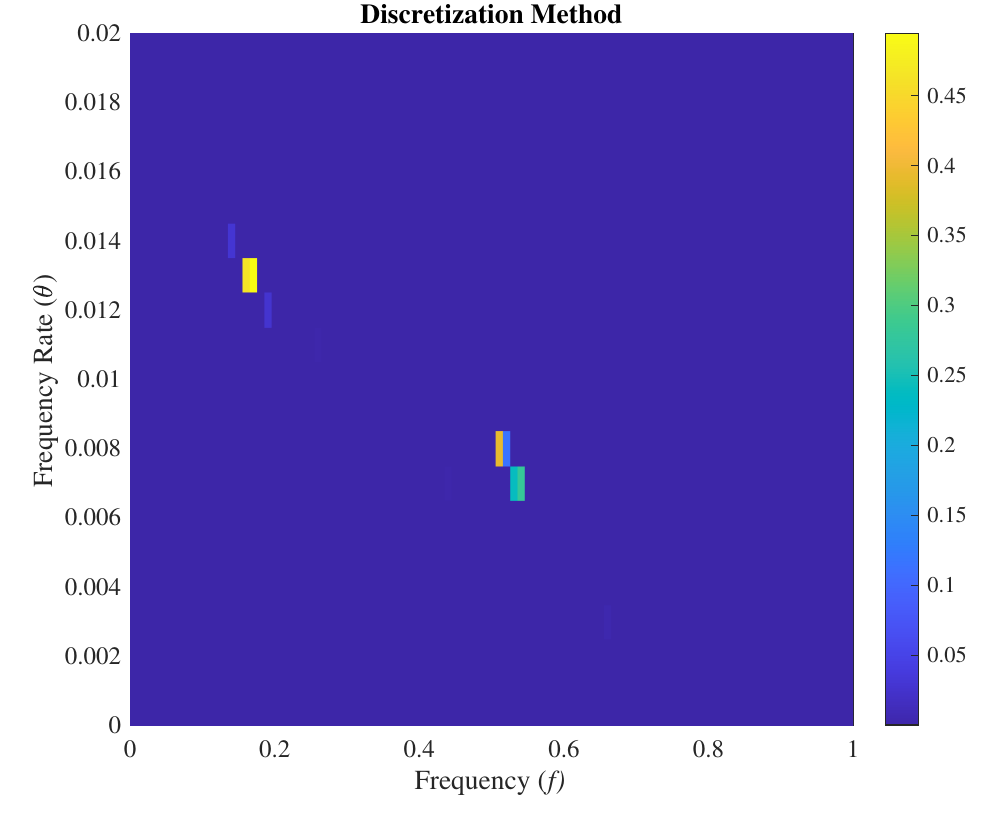}
   \caption{}
   \end{subfigure} 
   \vspace{3mm}
   \\
    \begin{subfigure}[b]{.49\columnwidth}
    \includegraphics[width=\columnwidth, height=1.5in]{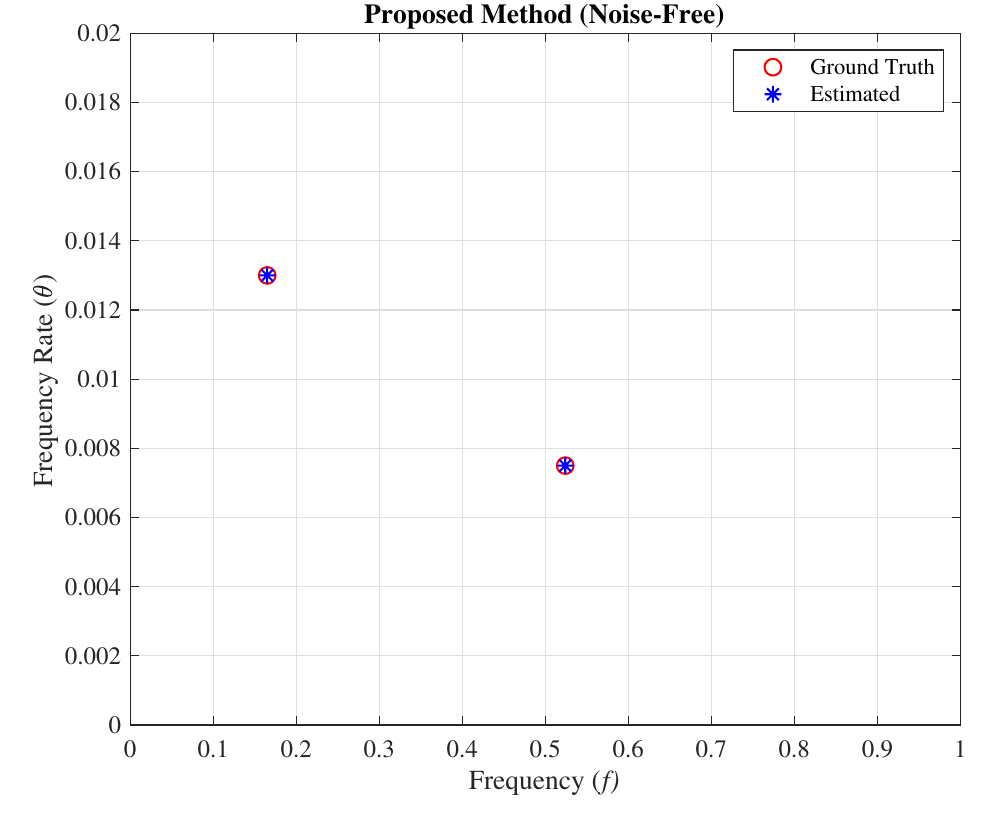}
         \caption{}
   \end{subfigure} 
    \begin{subfigure}[b]{.49\columnwidth}
    \includegraphics[width=\columnwidth, height=1.5in]{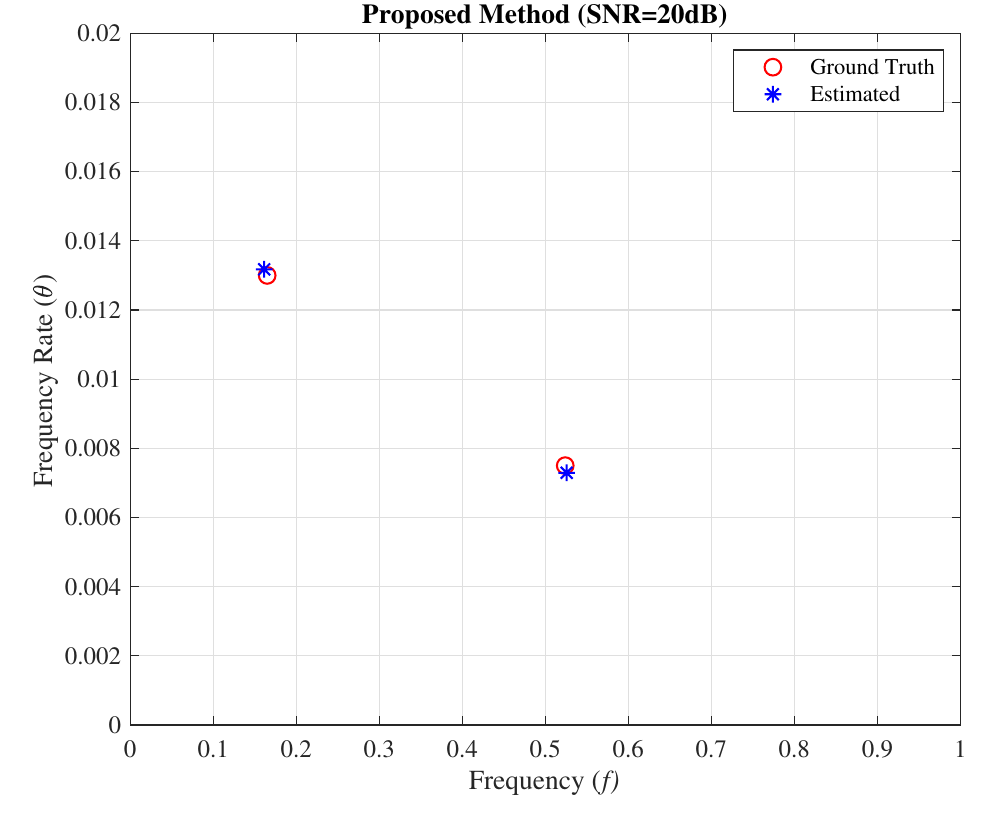}
      \caption{}
   \end{subfigure} 
\caption{The recovered continuous-valued chirp parameters in an illustrative example with $N = 25$ and $K = 2$. (a) Ground truth; (b) basis pursuit; (c) proposed approach; (d) proposed approach under noise.}
\end{figure}
\begin{figure}[htb]
\centerline{\includegraphics[width=1\linewidth]{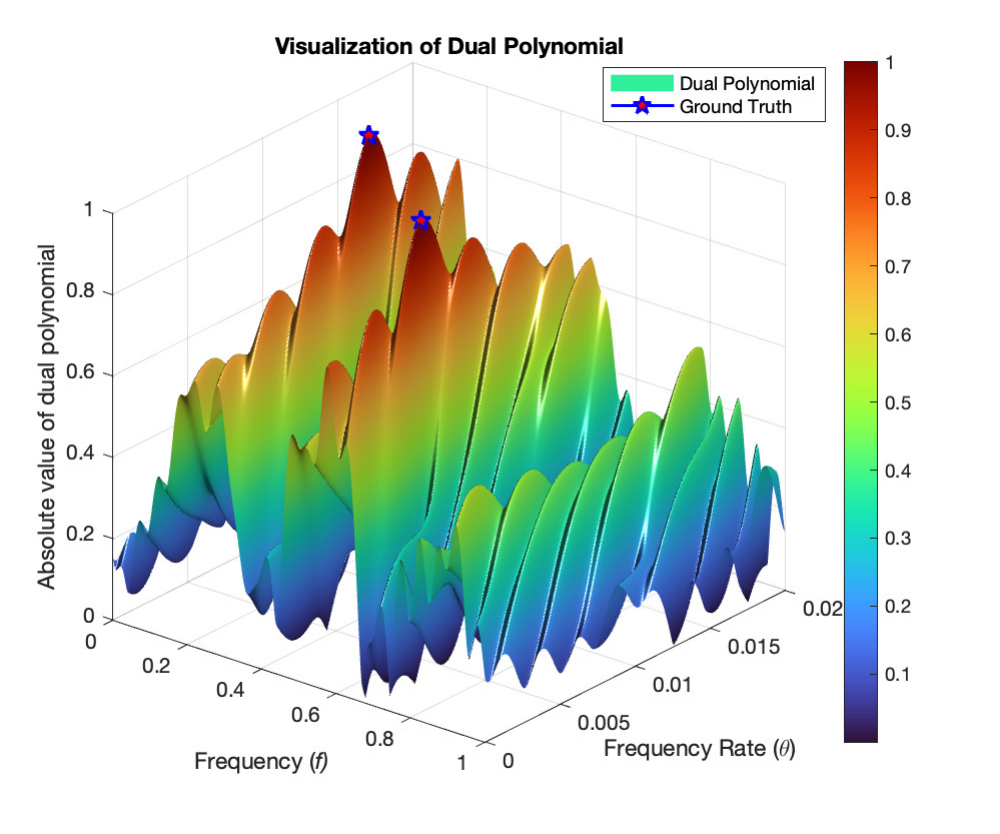}}
\caption{Localization of chirp parameters with the dual polynomial.}
\label{fig}
\end{figure}
\vspace{-3mm}
\section{Conclusion}
In this paper, we have taken a first step toward utilizing the atomic norm minimization framework for estimating the continuous parameters of linear chirp signals. By solving a constrained two-dimensional atomic norm minimization problem, we can recover the chirp parameters in the continuous domain without discretization. A theoretical analysis of the proposed approach is currently under preparation and will be presented in a future paper.

\bibliographystyle{IEEEbib}
\footnotesize
\bibliography{refs}

\begin{thebibliography}{10}

\bibitem{spread-specturm}
A.~Springer, W.~Gugler, M.~Huemer, L.~Reindl, C.~Ruppel, and R.~Weigel,
\newblock ``Spread spectrum communications using chirp signals,''
\newblock in {\em IEEE/AFCEA EUROCOMM 2000. Information Systems for Enhanced Public Safety and Security}, May 2000, pp. 166--170.

\bibitem{radarbook}
M.~Skolnik,
\newblock {\em Radar Handbook},
\newblock New York:McGraw-Hill, January 1990.

\bibitem{julian}
J.~Neri, P.~Depalle, and R.~Badeau,
\newblock ``Damped chirp mixture estimation via nonlinear {B}ayesian regression,''
\newblock in {\em Proceedings of the 24th International Conference on Digital Audio Effects}, Vienna, Austria, September 2021, pp. 65--72.

\bibitem{djuric}
P.~Djuric and S.~Kay,
\newblock ``Parameter estimation of chirp signals,''
\newblock {\em IEEE Trans. Acoust., Speech, Signal Processing}, vol. 38, pp. 2118--2126, 1990.

\bibitem{doweck-1}
Y.~Doweck, A.~Amar, and I.~Cohen,
\newblock ``Fundamental initial frequency and frequency rate estimation of random-amplitude harmonic chirps,''
\newblock {\em IEEE Trans. Signal Processing}, vol. 63, no. 23, pp. 6213--6228, 2015.

\bibitem{volcker}
B.~Volcker and B.~Ottersten,
\newblock ``Chirp parameter estimation from a sample covariance matrix,''
\newblock {\em IEEE Trans. Signal Processing}, vol. 49, no. 3, pp. 603--612, 2001.

\bibitem{sward}
J.~Sward, J.~Brynolfsson, A.~Jakobsson, and M.~Hansson-Sandsten,
\newblock ``Sparse semi-parametric estimation of harmonic chirp signals,''
\newblock {\em IEEE Trans. Signal Processing}, vol. 64, no. 7, pp. 1798--1807, 2016.

\bibitem{tf-atomic}
T.~Kusano, K.~Yatable, and Y.~Oikawa,
\newblock ``Sparse time-frequency representation via atomic norm minimization,''
\newblock in {\em Proceedings of IEEE International Conference on Acoustics, Speech and Signal Processing (ICASSP)}, Toronto, ON, Canada, June 2021, pp. 5075--5079.

\bibitem{tang2013compressed}
G.~Tang, B.~Bhaskar, P.~Shah, and B.~Recht,
\newblock ``Compressed sensing off the grid,''
\newblock {\em IEEE Trans. Information Theory}, vol. 59, no. 11, pp. 7465--7490, 2013.

\bibitem{chi-2d}
Y.~Chi and Y.~Chen,
\newblock ``Compressive two-dimensional complex exponentials from modulations with unknown waveforms,''
\newblock {\em IEEE Trans. Signal Processing}, vol. 63, no. 4, pp. 1030--1042, 2015.

\bibitem{chandrasekaran2012convex}
V.~Chandrasekaran, B.~Recht, P.~Parrilo, and A.~Willsky,
\newblock ``The convex geometry of linear inverse problems,''
\newblock {\em Foundations of Computational Mathematics}, vol. 12, no. 6, pp. 805--849, 2012.

\bibitem{dyang}
D.~Yang, G.~Tang, and M.~Wakin,
\newblock ``Super-resolution of complex exponentials from modulations with unknown waveforms,''
\newblock {\em IEEE Trans. Information Theory}, vol. 62, no. 10, pp. 5809--5830, 2016.

\bibitem{mishra}
K.~V. Mishra, M.~Cho, A.~Kruger, and W.~Xu,
\newblock ``Spectral super-resolution with prior knowledge,''
\newblock {\em IEEE Trans. Signal Processing}, vol. 63, no. 20, pp. 5342--5357, 2015.

\bibitem{yangzai-1}
Z.~Yang and L.~Xie,
\newblock ``Frequency-selective {V}andermonde decomposition of {T}oeplitz matrices with applications,''
\newblock {\em Signal Processing}, vol. 142, pp. 157--167, 2018.

\bibitem{li-yc}
Y.~Li, X.~Wang, and Z.~Ding,
\newblock ``Multidimensional spectral super-resolution with prior knowledge with application to high mobility channel estimation,''
\newblock {\em IEEE Journal of Selected Areas in Communications}, vol. 38, no. 12, pp. 2836--2852, 2020.

\bibitem{grant2008cvx}
M.~Grant, S.~Boyd, and Y.~Ye,
\newblock ``C{VX}: Matlab software for disciplined convex programming,'' 2008.

\bibitem{zhe-zhang-decouple}
Z.~Zhang, Y.~Wang, and Z.~Tian,
\newblock ``Efficient two-dimensional line spectrum estimation based on decoupled atomic norm minimization,''
\newblock {\em Signal Processing}, vol. 163, pp. 95--106, 2019.

\bibitem{feng-decouple}
F.~Xi, S.~Chen, and Z.~Liu,
\newblock ``Super-resolution delay-{D}oppler estmation for sub-{N}yquist radar via atomic norm minimization,''
\newblock in {\em Proceedings of IEEE International Conference on Acoustics, Speech and Signal Processing (ICASSP)}, New Orleans, LA, USA, March 2017, pp. 4326--4330.

\end{thebibliography}

\end{document}